\documentclass[twocolumn, conference]{IEEEtran}

\usepackage[style=ieee,backend=bibtex,sortcites=true,sorting=nyt,hyperref=false,backref=false]{biblatex}
\usepackage{microtype}

\usepackage[T1]{fontenc}
\usepackage[utf8]{inputenc}
\usepackage[english]{babel}
\usepackage{graphicx}
\usepackage[font=small,justification=centering,labelsep=period]{caption}
\usepackage{ifthen}
\usepackage{amssymb}
\usepackage{xcolor}
\usepackage{tabularx}
\usepackage{tabu}
\usepackage{url}
\usepackage{amsmath,amssymb}
\usepackage{multirow}
\usepackage{flushend}

\usepackage{hyperref}

\usepackage{booktabs}

\newboolean{showcomments}
\setboolean{showcomments}{true}
\ifthenelse{\boolean{showcomments}}
{ \newcommand{\mynote}[3]{
    \fbox{\bfseries\sffamily\scriptsize#1}
    {\small$\blacktriangleright$\textsf{\emph{\color{#3}{#2}}}$\blacktriangleleft$}}}
{ \newcommand{\mynote}[3]{}}

\graphicspath{{figures/}}
\bibliography{bib-sports}

\begin{document}

\title{Judging the Judges: \\ A General Framework for Evaluating the \\ Performance of International Sports Judges}

\author{
\IEEEauthorblockN{Sandro Heiniger}
\IEEEauthorblockA{Universität St.Gallen \\ Switzerland \\ sandro.heiniger@unisg.ch}
\and
\IEEEauthorblockN{Hugues Mercier}
\IEEEauthorblockA{Universit\'{e} de Neuch\^{a}tel \\ Switzerland \\ hugues.mercier@unine.ch}
}

\maketitle

\begin{abstract}

The monitoring of judges in sports is an important topic due to the media exposure of
international sporting events and the huge monetary sums that directly depend on the outcomes
of competitions. We present a method to assess the accuracy of sports judges applicable to all
sports where panels of judges evaluate athletic performances on a finite scale.  
We analyze judging scores from eight different sports with comparable judging systems: artistic
swimming, diving, dressage, figure skating, freestyle skiing, freestyle snowboard, gymnastics and
ski jumping. We identify, for each aforementioned sport, a general and accurate pattern of the
intrinsic judging error variability as a function of the performance level of the athlete. With the
notable exception of dressage, this intrinsic judging inaccuracy is heteroscedastic and can be
approximated by a concave quadratic curve, indicating increased consensus among judges towards the
best athletes. 
Using this observation, we can evaluate the performance of judges compared to their peers, and
distinguish erratic from precise judges and potential cheating from unintentional misjudging. Our
analysis also reveals valuable insights about the judging practices of the sports under
consideration, including a systemic problem in dressage where judges disagree on what
constitutes a good performance.

\end{abstract}

\textbf{Keywords:} Sports judges, quantifying accuracy, judging panels, heteroscedasticity, general framework.

\section{Introduction}\label{sec:Het_Introduction}

The essence of sporting competitions is that the outcome is not decided in advance, nor affected by events outside of what happens on the field of play. In many sports, judging decisions can make the difference between victory and defeat. Match fixing, bias and misjudgments by officials and referees are highly damaging to the reputation and bottom line of competitive sports. Honest athletes, coaches, fans, gamblers, officials and sponsors wish for accurate and fair judges. 

With the ever-expanding commercialization and media exposure of sporting events, judging decisions can bring fame and fortune for the winners and lifetime disappointment for the losers. These decisions can a have significant economic impact on the athletes' livelihood through corporate sponsorships and country bonuses. As one of countless examples, Singapore promises \$1,000,000 to individual gold medalists at the Olympic Games \cite{Singapore}.

Judges and referees are subject to many biases. The most important is national bias, which appears in many sports \cite{Ansorge:1988,Campbell:1996,Popovic:2000,Zitzewitz:2006, Emerson:2009, Leskovsek:2012,Zitzewitz:2014,Sandberg:2018,HM2018:NationalBias}, and results in judges awarding higher marks to athletes of their own nationality. Besides these biases, another incentive to monitor sports judges is the exponential increase of sports betting, especially online, which is now legal in many countries. The May 2018 Supreme Court ruling in favor of New Jersey against the 1992 federal ban on sports betting has led several US states to legalize sports gambling, with many other states actively working on its legalization. It has never been easier to legally bet on the outcomes of sporting events. Of course, sports betting as well as match fixing and corruption by judges and referees have gone hand in hand for ages, as recently and shamefully illustrated by the 2005 Bundesliga\footnote{\url{https://en.wikipedia.org/wiki/Bundesliga_scandal_(2005)}}, 2005 Máfia do Apito\footnote{\url{https://en.wikipedia.org/wiki/Brazilian_football_match-fixing_scandal}} and 2007 NBA scandals\footnote{\url{https://en.wikipedia.org/wiki/2007_NBA_betting_scandal}}. 

All these factors have led to increased scrutiny of judges and referees. \textcite{Price:2010} quote former NBA Commissioner David Stern claiming that NBA referees "are the most ranked, rated, reviewed, statistically analyzed and mentored group of employees of any company in any place in the world". However, despite many such claims, there is little public disclosure from sports leagues, federations and governing bodies on how they monitor their judges, and most are reluctant to share data and processes to outsiders. For instance, following the 2002 Winter Olympics figure skating scandal\footnote{\url{https://en.wikipedia.org/wiki/2002_Winter_Olympics_figure_skating_scandal}}, the International Skating Union (ISU) reformed its scoring principles, and anonymized marks given by judges. Despite the intent of reducing collusion and corruption, anonymization made it more difficult for third parties to monitor judges. In fact, \textcite{Zitzewitz:2014} showed that national bias and vote trading increased after the rule changes. Following dubious judging at the 2014 Sochi Winter Olympics, the ISU backtracked and removed judge anonymity.

The last reason to monitor judges and referees, perhaps not as scandalous but more important on a day to day basis, is the simple fact that some judges are more accurate than others at judging. Assessing the skill level of sports judges objectively is a difficult endeavor and there is little literature on the topic. This is the main objective of this work.  

\subsection{Our contributions}

We present a general framework to evaluate the performance of international sports judges, applicable to all sports where panels of judges evaluate athletes on a finite scale.  For all these sports, judge evaluations decide the entirety, or a large fraction, of the outcome of the performances. As opposed to professional sports leagues, who have well-compensated and unionized referees, most judges in the sports we target are unpaid volunteers. They receive a small stipend covering their travel expenses when they officiate, but otherwise have other jobs and do judge duties by passion for their sport. Although their training, selection and monitoring vary per sport, they are more susceptible to favoritism\footnote{\url{https://www.washingtonpost.com/sports/olympics/two-chinese-figure-skating-judges-suspended-for-showing-favoritism-at-games/2018/06/21/8c0dc3e6-7557-11e8-b4b7-308400242c2e_story.html?noredirect=on&utm_term=.294cf291bd2f}}, collusion\footnote{\url{https://globalnews.ca/news/1139349/figure-skating-scoring-lends-itself-to-scandal-and-getting-worse-expert/}}, bribery\footnote{\url{https://www.theguardian.com/sport/2016/aug/01/rio-2016-olympics-boxing-corruption-allegations}}, and more simply but as importantly, lack of competence\footnote{\url{https://slate.com/culture/2016/08/are-olympic-boxing-judges-incompetent-corrupt-or-both.html}}.

We test and confirm our framework using data from artistic swimming\footnote{Formerly known as synchronized swimming.}, diving, dressage, figure skating, freestyle skiing (aerials), freestyle snowboard (halfpipe, slopestyle), gymnastics and ski jumping international competitions. All these sports are part of the Olympic family, and their federations derive a significant portion of their operating budgets from broadcasting and marketing rights of Olympic Games redistributed by the IOC~\cite{IOC:2019}. Furthermore, artistic swimming, dressage and rhythmic gymnastics are less popular worldwide, have the reputation of being more subjective, and are often mentioned when discussing sports that should removed from the Olympics. They thus have strong economic incentives to have objective competitions. 

Our main observation is that for all these sports except dressage, the standard deviation of the judging error is heteroscedastic, and we can model it accurately using a concave quadratic equation: judges are more precise when evaluating outstanding or atrocious performances than when evaluating mediocre ones. We can use this observation to quantify the long-term accuracy of judges against their peers. We provide evidence that the implemented scoring systems generally lead to objective judging. The exception is dressage, where judges increasingly disagree with each other as the performance quality improves, which implies a lack of objectivity compared to the other sports we analyze.

This is the third in a series of three articles on sports judging, extending our initial work on gymnastics. In the first article~\cite{MH2018:gymnastics}, we model the intrinsic judging error of international gymnastics judges as a function of the performance level of the gymnasts using heteroscedastic random variables. We then develop a marking score to quantify the accuracy of international gymnastics judges. In the second article~\cite{HM2018:NationalBias},
we leverage the heteroscedasticity of the judging error of gymnastics judges to improve the assessment of national bias in gymnastics.

\section{Related work on judging skill and heteroscedasticity}

The vast majority of research assessing judging skill in sports focuses on consensus and consistency within groups. In 1979, \textcite{Shrout:1979} introduced the idea of intra-class correlation. This technique was used to evaluate judging in figure skating \cite{Looney:2004}, artistic gymnastics \cite{Bucar:2014, Leskovsek:2012, Atikovic:2009} and rhythmic gymnastics \cite{Leandro:2017}. The dependence between the variability of the judging error and  performance quality had never been properly studied until our work in gymnastics~\cite{MH2018:gymnastics}, 
although it was observed in prior work. \textcite{Atikovic:2009} notice a variation of the marks deviation by apparatus in artistic gymnastics. \textcite{Leandro:2017} observe that the deviation of scores is smaller for the best athletes in rhythmic gymnastics, and \textcite{Looney:2004} notices the same thing in figure skating. 

Even though heteroscedasticity is often under-reported in scientific research \cite{Cleasby:2011}, it is a well known property and appears in countless fields such as biology \cite{SV:2015}, chemistry \cite{Rocke:1995}, economics \cite{Okun:1971,Pagan:1983}, finance \cite{SS1990,Nelson:1991} and sports science \cite{NA:1997}. In most cases, heteroscedasticity arises as a scale effect, i.e., the variance is linked to the size of the measurement. This is not the case in our work, where the variability of the judging error depends on the quality of the performance~--~a~completely different source of variability than the scale effect. 

Another particularity of our work is that instead of having a single observer making measurements, a set of observers, i.e., a panel of sports judges, independently observe and measure a common set of performances. Heteroscedasticity in a similar context was observed in judicial decision making \cite{Collins:2008}, and is implicit in the data of wine tasting scores \cite{Cicchetti:2004b}. Thus, as opposed to prior work, our goal is not only to model a heteroscedastic random variable accurately, but to assess the accuracy of the observers. This, to the best of our knowledge, has never been systematically attempted before our work in gymnastics.

\section{Judging systems and dataset}
We analyze eight sports with comparable judging systems: artistic swimming, diving (including  high-diving), dressage (Grand Prix, Grand Prix Special \& Grand Prix Freestyle), figure skating,  freestyle skiing (aerials), freestyle snowboard (halfpipe, slopestyle), gymnastics (acrobatic, aerobic, artistic, rhythmic, trampoline) and ski jumping. For all these sports, it is impossible to evaluate performances in an automated fashion, and a panel of judges evaluates the athletes. Each judge in the panel reports an individual mark within a closed finite range following predefined judging guidelines. Although the implementation details such as the precise judging guidelines, the marking range and the step size vary, all the judging systems incorporate
\begin{enumerate}
	\item An execution evaluation of performance components;
	\item Penalty deductions for mistakes.
\end{enumerate}
The judging guidelines of each sport are the embodiment of the performance quality, mapped to a closed finite nominal mark. In particular, they embed the concept of perfection: a component or routine whose mark is the maximum possible value is considered perfect.

Although the judging guidelines try to be as deterministic and accurate as possible, every reported mark inevitably remains a subjective approximation of the actual performance level. Moreover, live judging is a noisy process, and judges within a panel rarely agree with each other. There are well-documented psychological, physiological, environmental and organizational factors that make this process noisy, and the best judges are the ones that systematically minimize this noise. Every judging system hence accounts for this variability with an aggregation of the judging panel marks to ensure an accurate and fair evaluation of the athletes. This aggregation, which varies for each sport, is also used to discard outlier marks, and is the most effective way to decrease the influence of erratic, biased or cheating judges.

\begin{table}
	\centering 
	\begin{tabular}{lccc}
		\toprule
		& Size of &  Number of & Number of\\
		Sport & judging panel & performances & marks\\ 
		\midrule
		Aerials & 3 or 5 & 7'079 & 53'543\\
		Artistic swimming & 5, 6 or 7 & 3'882 & 42'576  \\
		Diving & 7 & 19'111 & 133'777\\ 
		Dressage &&&\\
		\quad $\cdot$ GP \& GP Special & 5 or 7 & 5'500 & 28'382\\
		\quad $\cdot$ GP Freestyle & 5 or 7 & 2'172 & 11'162\\
		Figure skating & 9  & 5'076 & 228'420 \\ 
		Freestyle snowboard &&& \\
		\quad $\cdot$ Halfpipe & 3, 4, 5 or 6 & 4'828 & 22'389\\ 
		\quad $\cdot$ Slopestyle & 3, 4, 5 or 6 & 2'524 & 12'374\\ 
		Gymnastics & & & \\
		\quad $\cdot$ Acrobatics & 6 or 8 & 756 & 4'870 \\
		\quad $\cdot$ Aerobics & 6 or 8 & 938 & 6'072 \\
		\quad $\cdot$ Artistics & 4, 6 or 7 & 11'940 & 78'696 \\
		\quad $\cdot$ Rhythmics & 4, 5 or 7 & 2'841 & 19'052 \\
		\quad $\cdot$ Trampoline & 4 or 5 & 1'986 & 9'654 \\
		Ski jumping & 5 & 12'381 & 61'905\\ 	
		
		\bottomrule
	\end{tabular} 
	\caption{Typical judging panel and size of our dataset per sport.}
	\label{tab:HET_Data}
\end{table}
Table \ref{tab:HET_Data} shows the typical judging panel and size of our dataset per sport. The number of marks counts every single reported mark by a judge in the dataset and depends on the number of performances or routines, the number of evaluated components per performance, and the size of the judging panel. The size of the judging panel ranges from three to nine judges and can vary within a sport depending on the stage or importance of the competition. 

The Fédération Internationale de Gymnastique (FIG)\footnote{www.gymnastics.sport} and Longines\footnote{www.longines.com} provided the data for all the gymnastics disciplines. It includes 21 continental and international competitions from the 2013--2016 Olympic cycle, ending with the 2016 Rio Olympic Games. We gathered data for all the other sports from publicly available sources at official federation or result websites\footnote{fis-ski.com/DB/ (aerials, halfpipe, ski jumping, slopestyle); www.omegatiming.com (diving); data.fei.org (dressage); www.isu.org (figure skating); www.swimming.org (artistic swimming); all retrieved September 1, 2017.}. The data only  comprises professional and international competitions. When available, the analysis includes all results of official World cup events, international championships and Olympic Games from January 2013 to August 2017. None of the sports has a gender-specific scoring system, thus we include men and women competitions in the dataset. 

Table~\ref{tab:HET_Data} excludes some of the gathered data to ensure comparability among the sports as follows. First, the analysis focuses on the execution and artistry components of performances, thus we exclude difficulty scores of technical elements from the sample. Acrobatic and aerobic gymnastics have separate marks for the execution and artistry components; we split the marks in our analysis, but do not distinguish between them because judges have the same judging behavior in both instances~\cite{MH2018:gymnastics}. In dressage, we only consider events at the Grand Prix level, which includes 'Grand Prix', 'Grand Prix Special' and 'Freestyle' competitions. Judges in figure skating and artistic swimming evaluate the execution of multiple components of a performance separately, which we group together in our analysis. Scores in aerials competitions consist of a 'Air', 'Form', and 'Landing' mark, although this granularity is not available for all competitions. We add the three marks together when analyzing total scores, and study the components separately when they are available.

\section{Methods}
\begin{table}
	\centering 
	\setlength{\tabcolsep}{1pt} 
	\begin{tabularx}{0.78\columnwidth}{cp{5mm}l} 
		\toprule
		$s_{p,j}$ && Mark from judge $j$ for performance $p$ \\
		$c_p$ && Control score of performance $p$ \\
		$\hat{e}_{p,j}$ && Judging error of judge $j$ for performance $p$ \\
		$\sigma_d(c)$ && Standard deviation of the judging error $\hat{e}_{p,j}$ for \\ && \qquad  discipline $d$ and performance level $c$ \\
		$\hat{\sigma}_d(c)$ && Approximate standard deviation of the judging error \\ && \qquad  for discipline $d$ as a function of the performance level $c$\\
		\bottomrule
	\end{tabularx} 
	\caption{Notation} 
	\label{tab:Notation} 
\end{table}
We perform the same analysis for every sport and discipline. Table \ref{tab:Notation} summarizes our notation. Let $s_{p,j}$ be the mark from judge $j$ for performance $p$. For each performance, we need a control score $c_p$ providing an objective measure of performance quality. We use the median panel score $c_p\triangleq\underset{j\text{ in panel}}{\text{median }}s_{p,j}$ in our analysis. The median is more robust than the mean or the trimmed mean against misjudgments and biased judges, and in the aggregate provides a good approximation of the true performance quality. In some sports such as gymnastics~\cite{MH2018:gymnastics}, more accurate control scores are derived using video analysis post competition, however these are not accessible for our analysis.

The difference $\hat{e}_{p,j}\triangleq s_{p,j} - c_p$ is the judging discrepancy of judge $j$ for performance $p$, which we use as a proxy for the judging error. For a given discipline $d$, we group the judging errors by control score $c$ and calculate the standard deviation $\sigma_d(c)$, quantifying how judges agree or disagree with each other for a given performance quality~$c$. We call $\sigma_d(c)$ the \emph{intrinsic discipline judging error variability}. We then approximate this variability as a function of performance quality with a polynomial of second degree $\hat{\sigma}_d(c)$ using a weighted least-squares quadratic regression.

\section{Results and Discussion}

\subsection{The general pattern of heteroscedasticity}
\begin{figure}
	\centering
	\includegraphics[width=1\columnwidth]{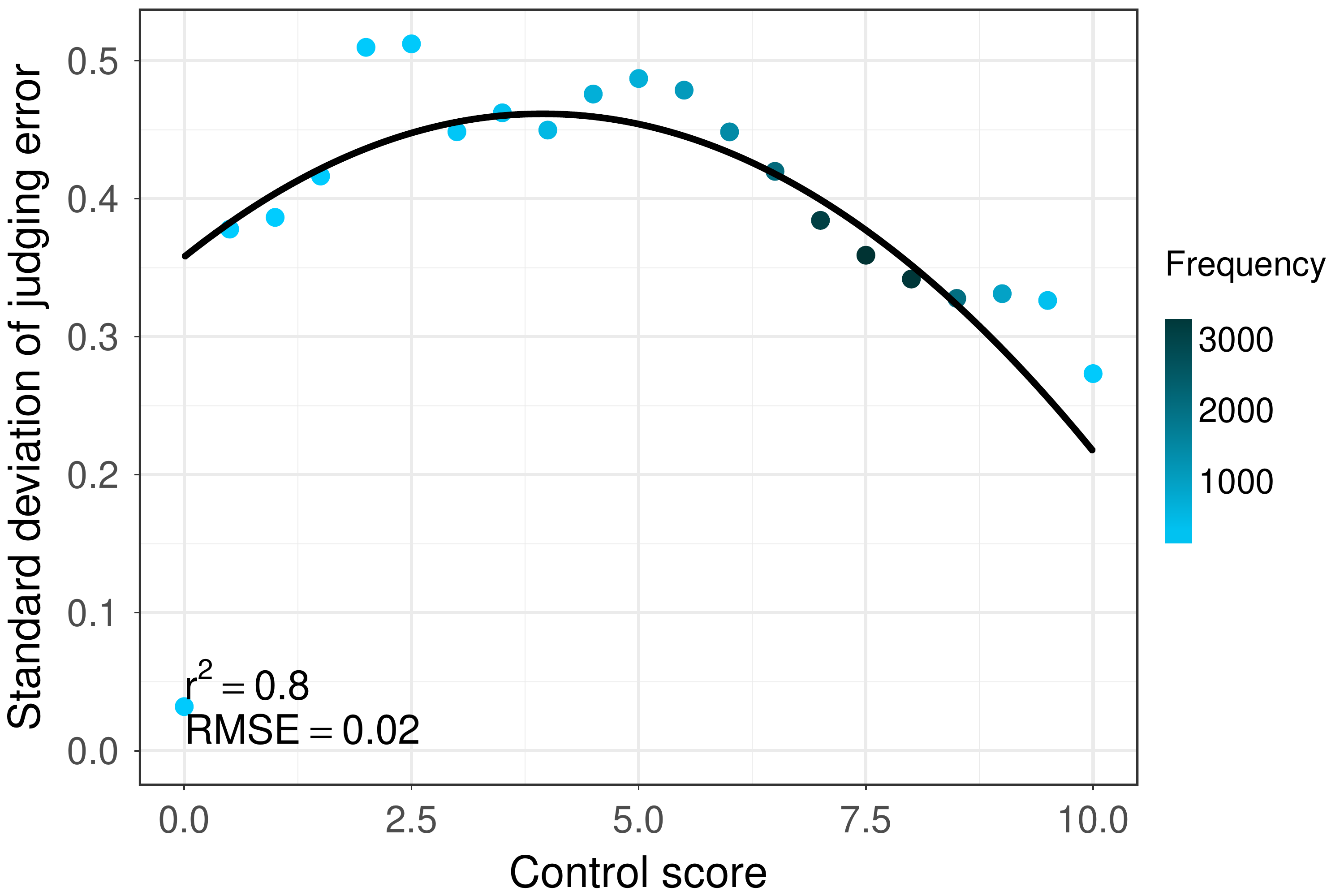}
	\caption{Standard deviation of judging marks versus performance quality in diving.}
	\label{fig:Res_Diving}
\end{figure}

\begin{figure}
	\centering
	\includegraphics[width=1\columnwidth]{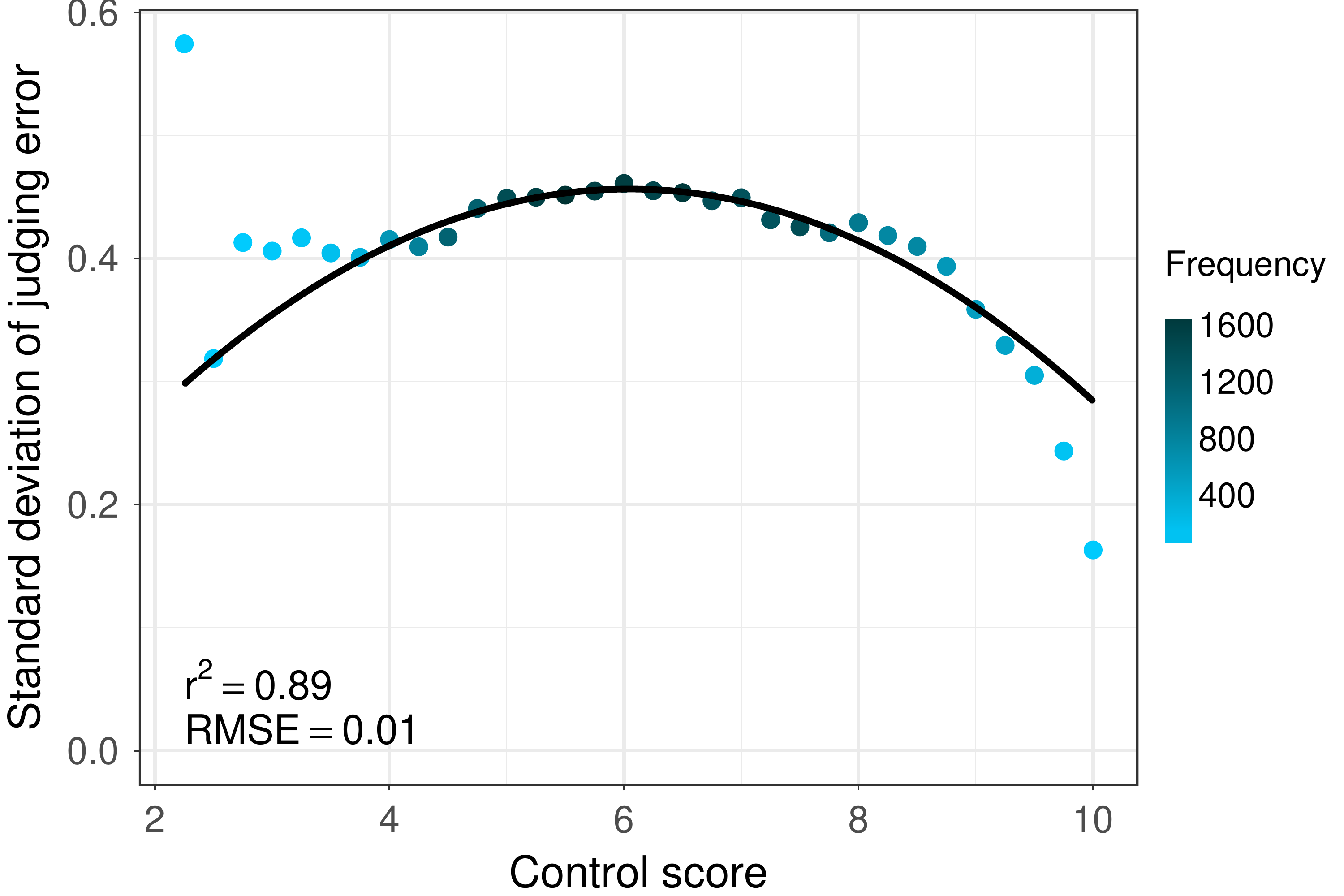}
	\caption{Standard deviation of judging marks versus performance quality in figure skating.}
	\label{fig:Res_Figureskating}
\end{figure}

\begin{figure}
	\centering
	\includegraphics[width=1\columnwidth]{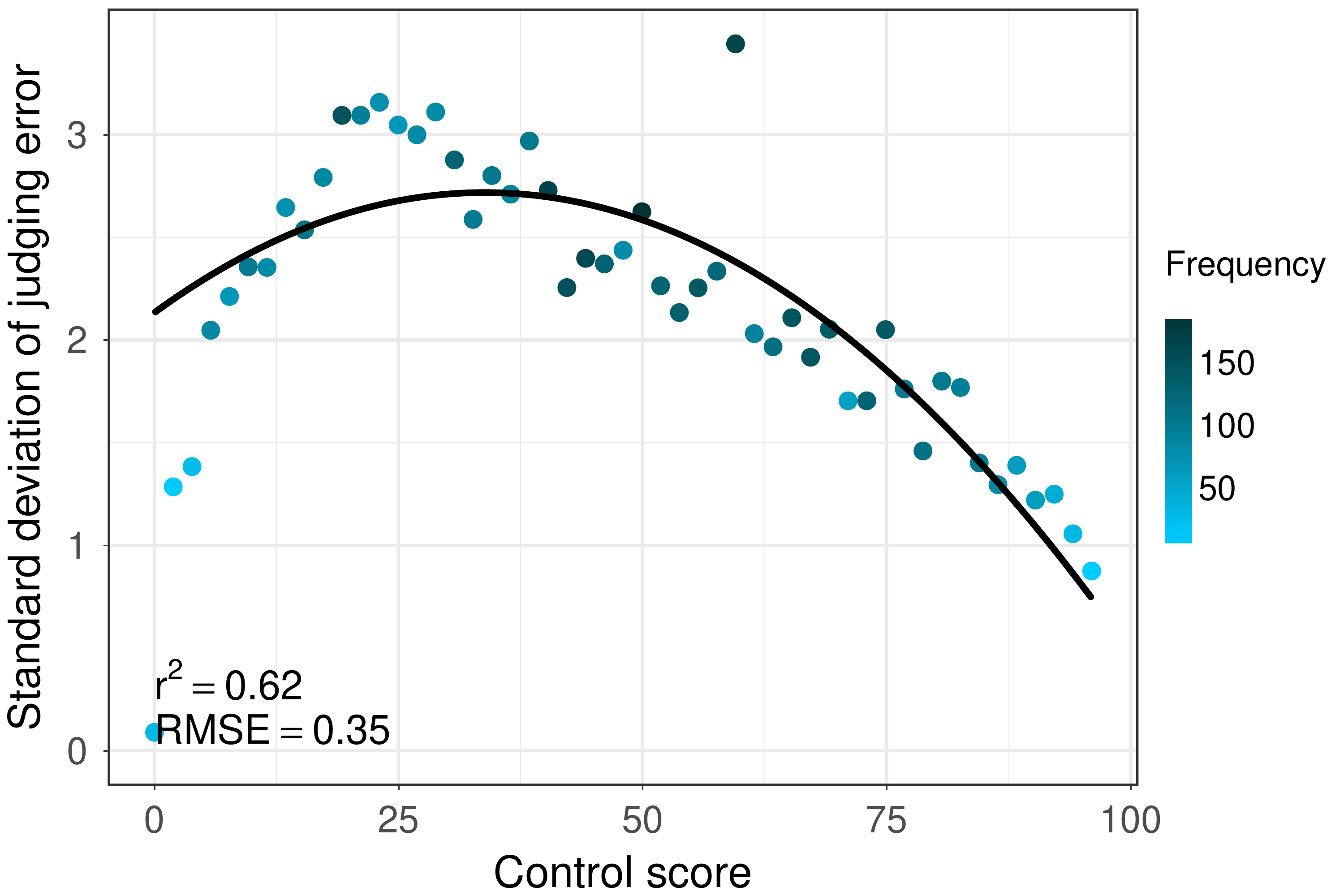}
	\caption{Standard deviation of judging marks versus performance quality in halfpipe.}
	\label{fig:Res_Halfpipe}
\end{figure}

\begin{figure}
	\centering
	\includegraphics[width=1\columnwidth]{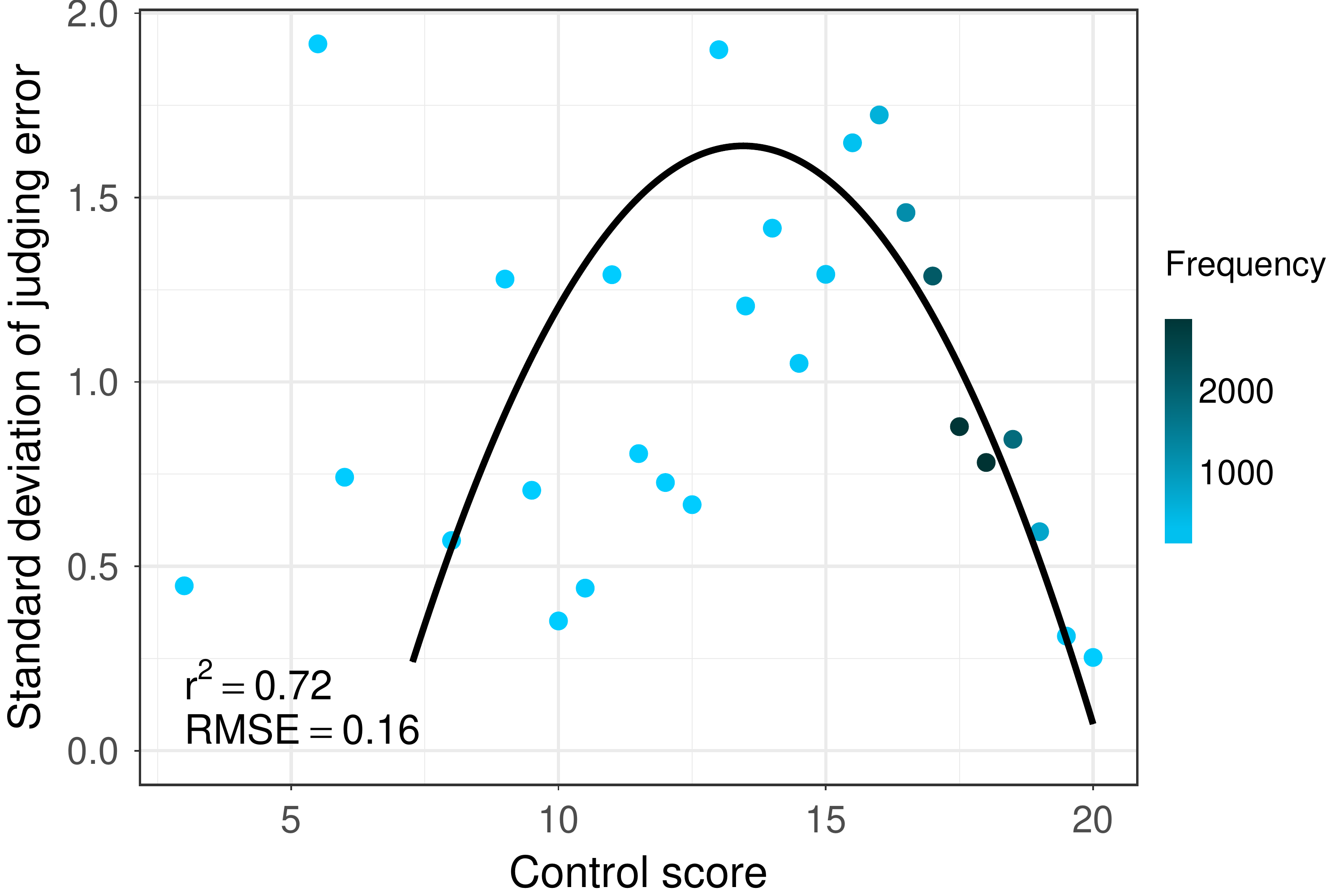}
	\caption{Standard deviation of judging marks versus performance quality in ski jumping.}
	\label{fig:Res_Skijumping}
\end{figure}

\begin{figure}
	\centering
	\includegraphics[width=1\columnwidth]{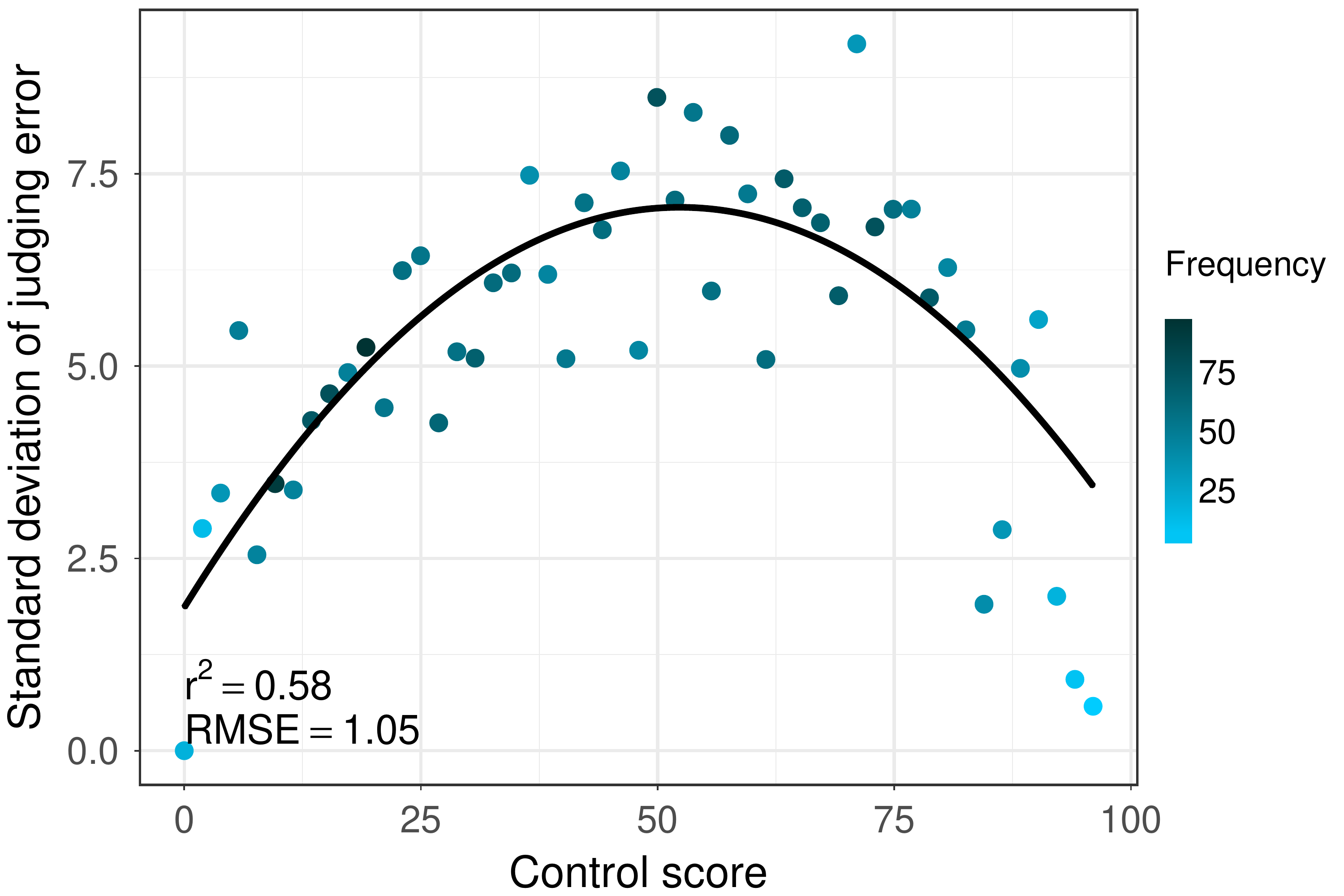}
	\caption{Standard deviation of judging marks versus performance quality in slopestyle.}
	\label{fig:Res_Slopestyle}
\end{figure}

\begin{figure}
	\centering
	\includegraphics[width=1\columnwidth]{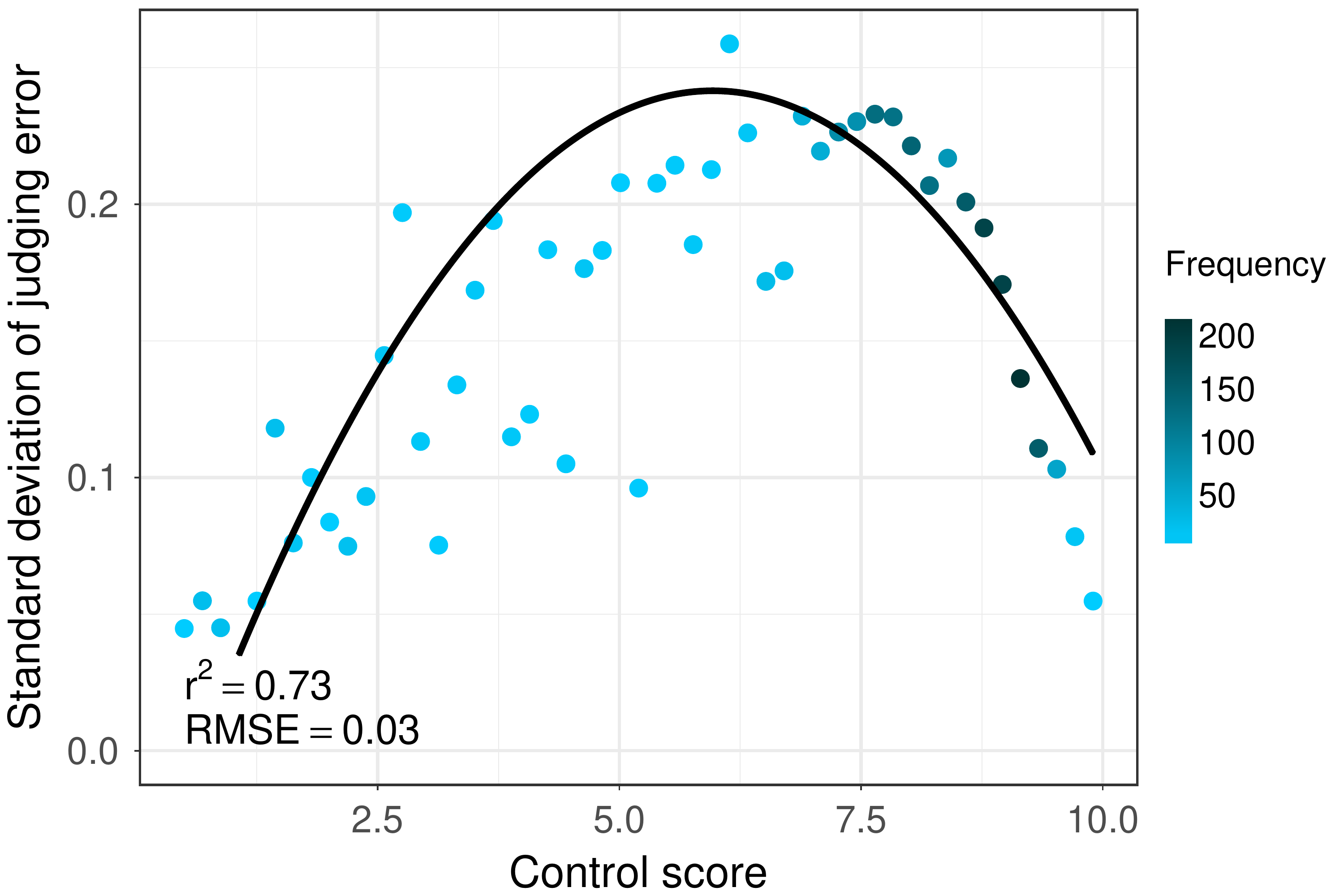}
	\caption{Standard deviation of judging marks versus performance quality in trampoline.}
	\label{fig:Res_Trampoline}
\end{figure}

\begin{figure}
	\centering
	\includegraphics[width=1\columnwidth]{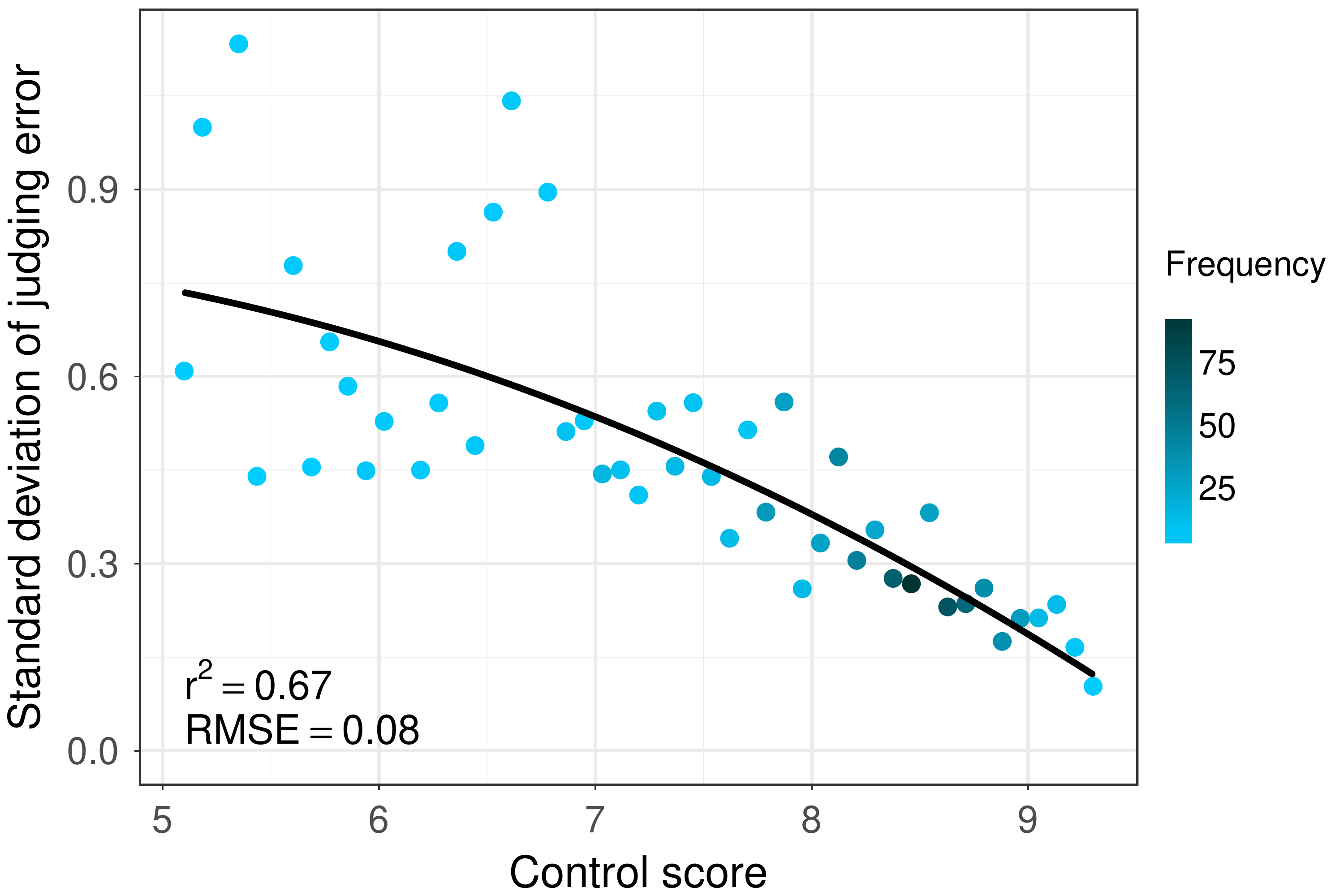}
	\caption{Standard deviation of judging marks versus performance quality in acrobatic gymnastics.}
	\label{fig:Res_Acrobatics}
\end{figure}

\begin{figure}
	\centering
	\includegraphics[width=1\columnwidth]{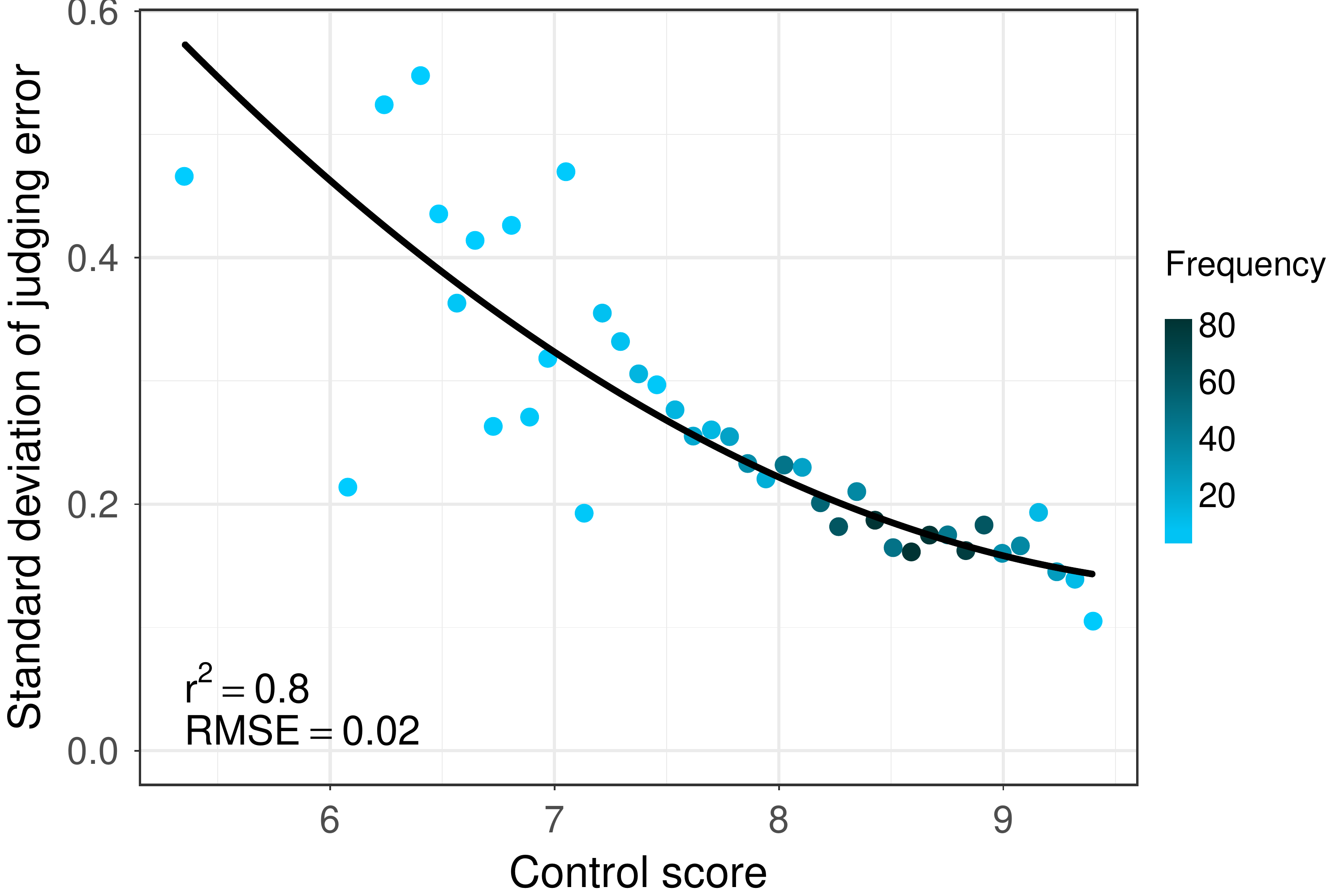}
	\caption{Standard deviation of judging marks versus performance quality in aerobic gymnastics.}
	\label{fig:Res_Aerobics}
\end{figure}

\begin{figure}
	\centering
	\includegraphics[width=1\columnwidth]{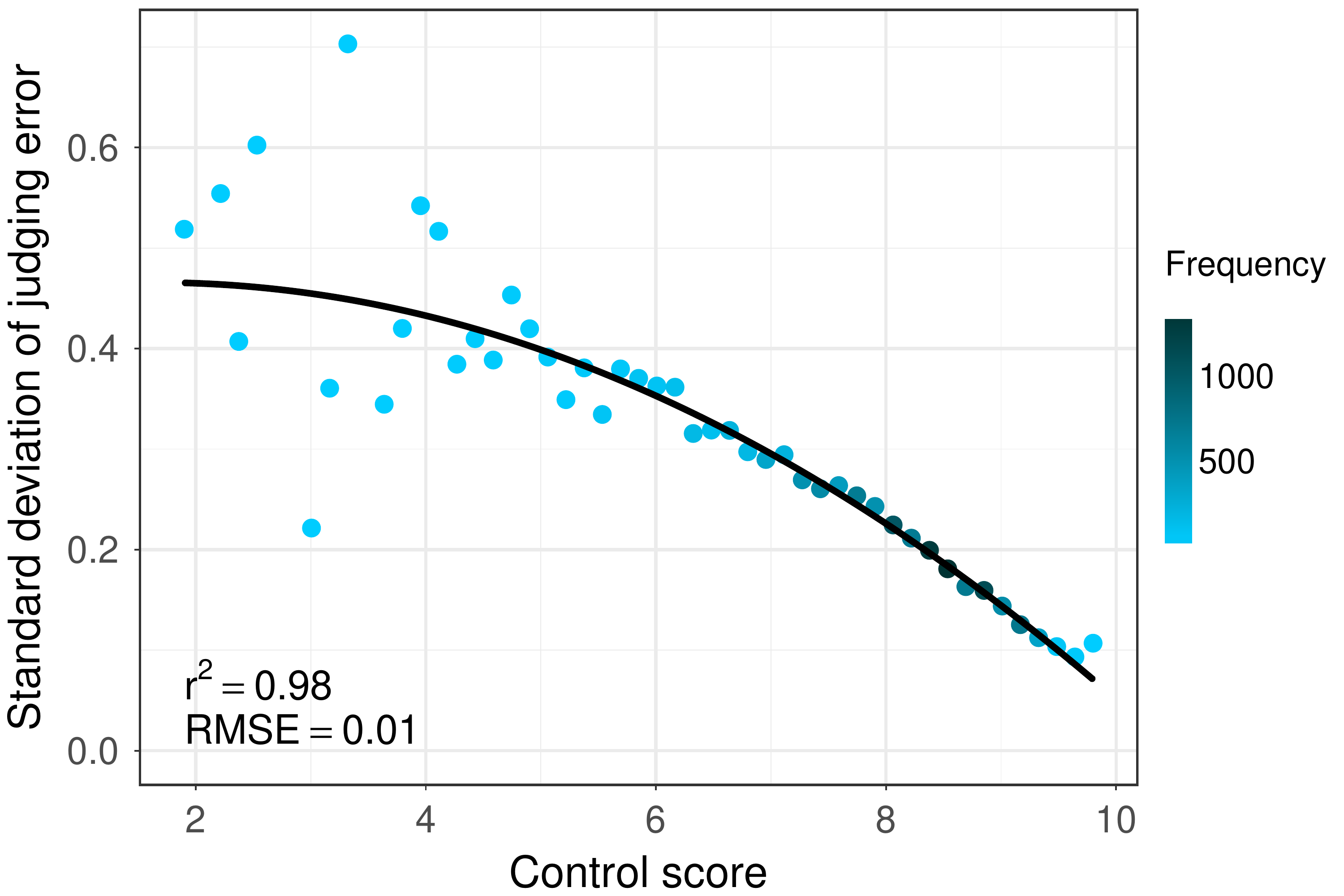}
	\caption{Standard deviation of judging marks versus performance quality in artistic gymnastics.}
	\label{fig:Res_Artistics}
\end{figure}

\begin{figure}
	\centering
	\includegraphics[width=1\columnwidth]{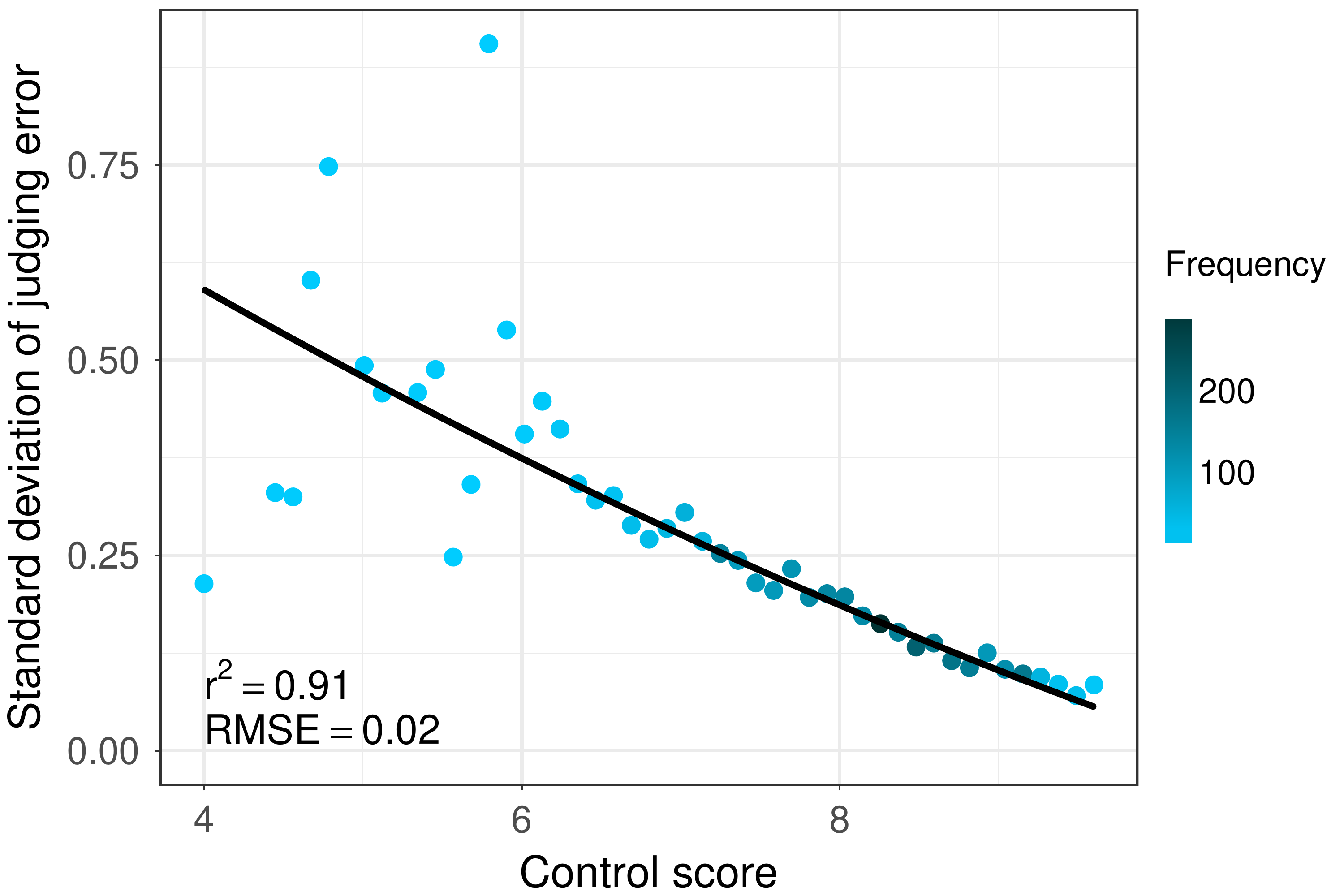}
	\caption{Standard deviation of judging marks versus performance quality in rhythmic gymnastics.}
	\label{fig:Res_Rhythmics}
\end{figure}

\begin{figure}
	\centering
	\includegraphics[width=1\columnwidth]{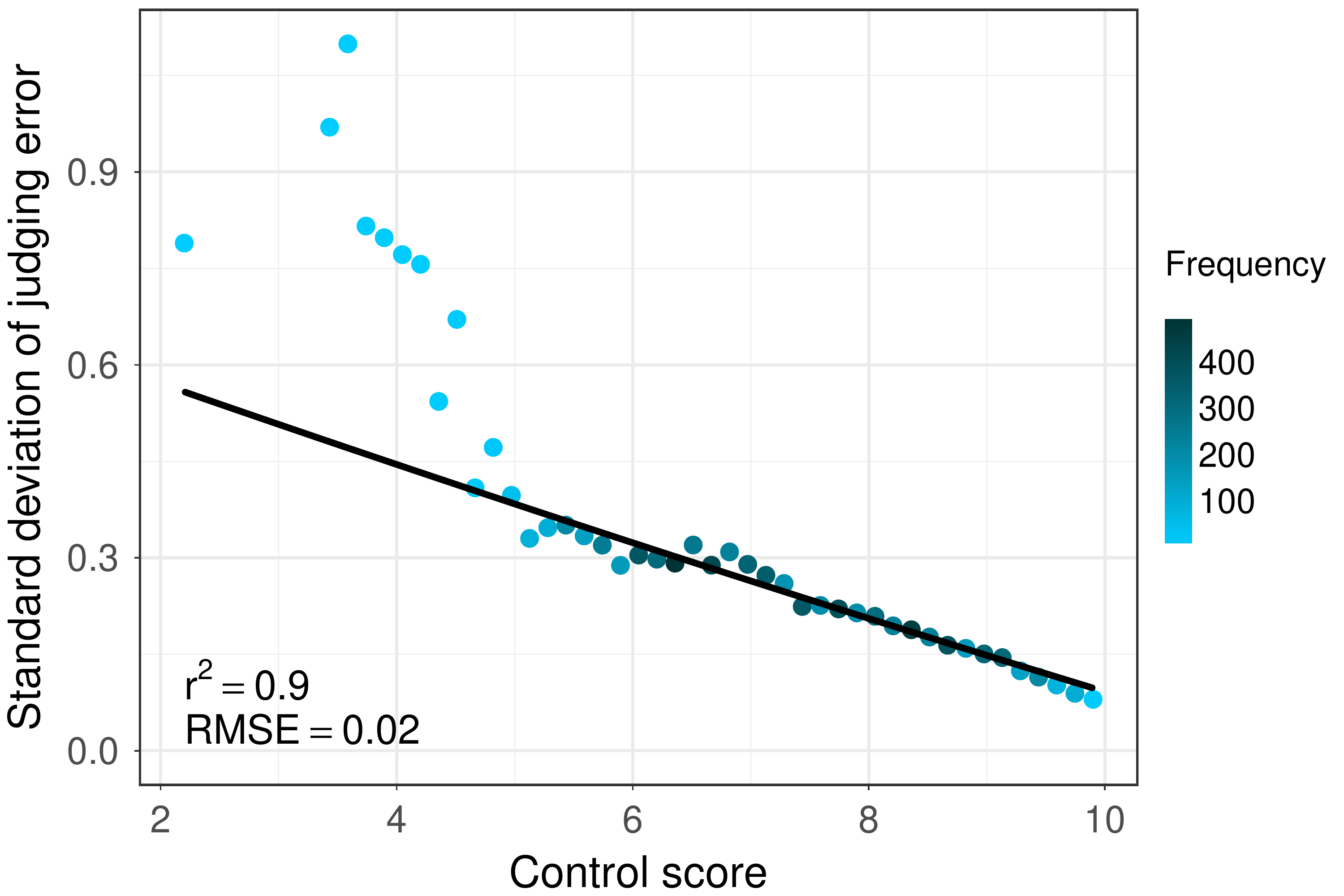}
	\caption{Standard deviation of judging marks versus performance quality in artistic swimming.}
	\label{fig:Res_Synchronisedswimming}
\end{figure}

\begin{figure}
	\centering
	\includegraphics[width=1\columnwidth]{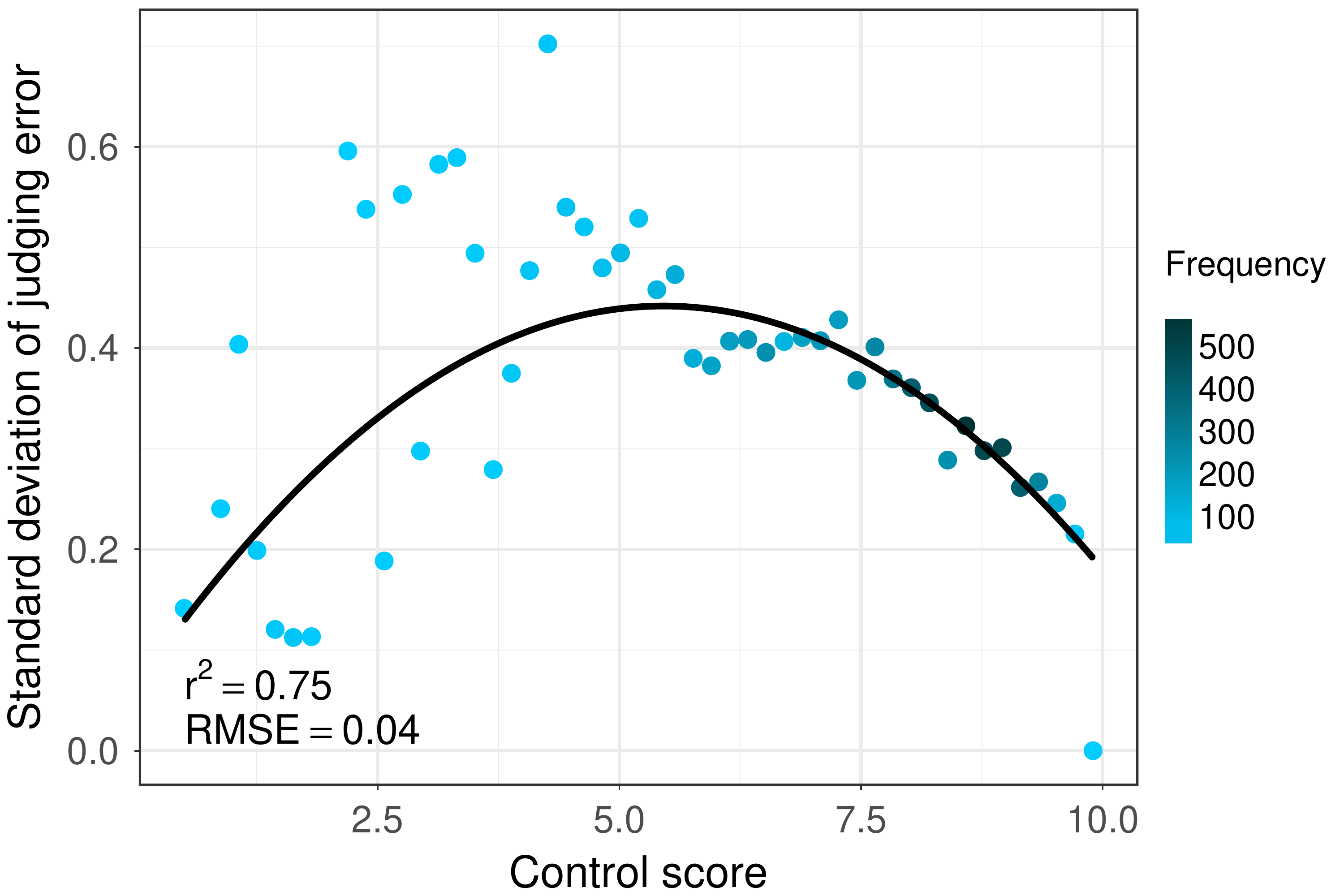}
	\caption{Standard deviation of judging marks versus performance quality in aerials (total scores).}
	\label{fig:Res_Aerials}
\end{figure}

\begin{figure}
	\centering
	\includegraphics[width=1\columnwidth]{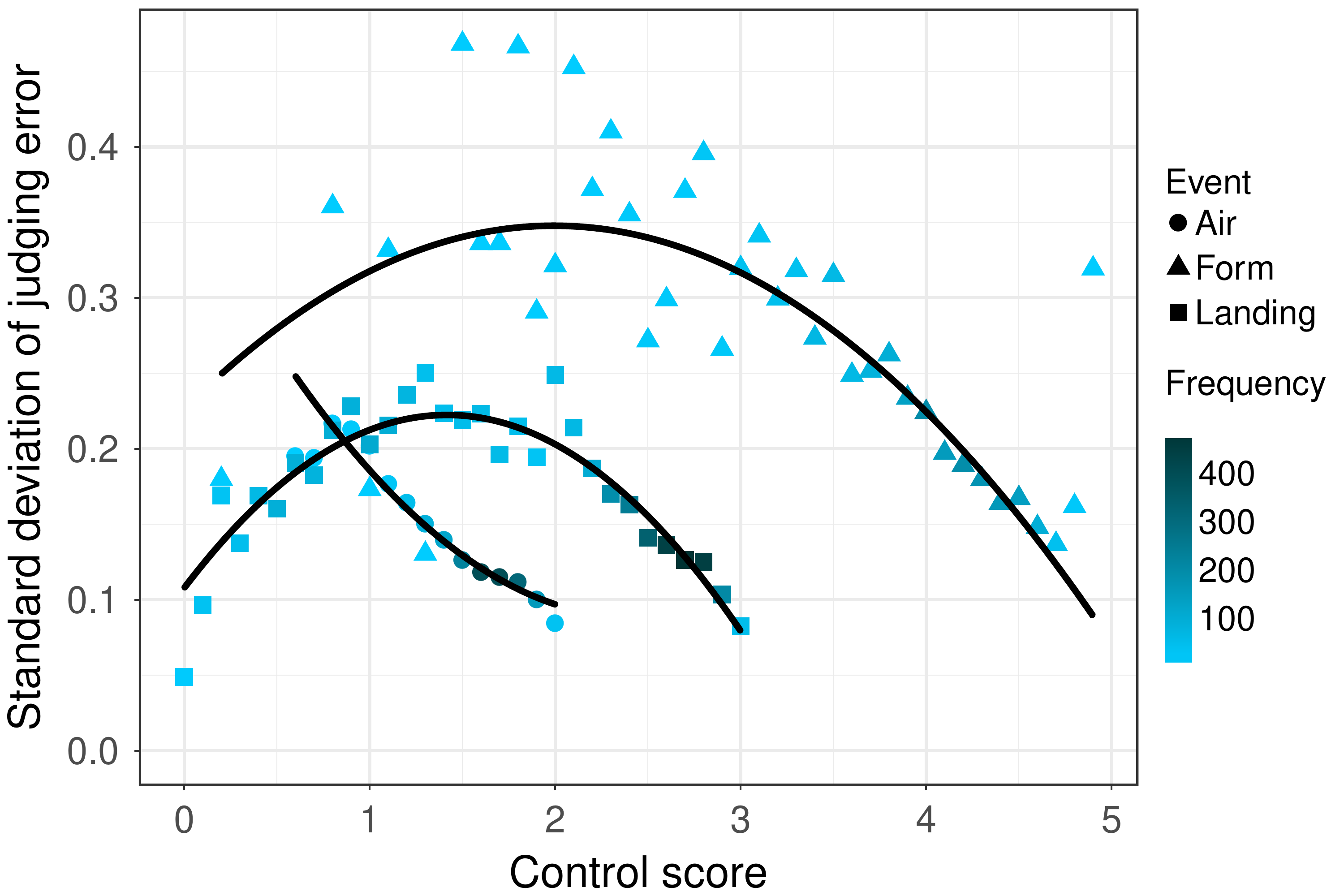}
	\caption{Standard deviation of judging marks versus performance quality in aerials (component scores).}
	\label{fig:Res_Aerialssplit}
\end{figure}

Figures \ref{fig:Res_Diving}-\ref{fig:Res_Dressage_split} show the standard deviation of the judging marks $\sigma_d(c)$ and the weighted least-squares quadratic regression curve $\hat{\sigma}_d(c)$ as a function of performance~$c$ for diving, figure skating, halfpipe, ski jumping, slopestyle, trampoline, acrobatic gymnastics, aerobic gymnastics, artistic gymnastics, rhythmic gymnastics, artistic swimming, aerials (total and component scores) and dressage (regular and freestyle to music events), respectively\footnote{Note that for some sports we aggregate close quality levels (control scores) to improve the visibility of the figures. We do the analysis without the aggregation.}. Each figure includes the scaleless weighted coefficient of determination $r^2$ quantifying the goodness of fit of the regression. The weighted $r^2$ are high, ranging from $0.24$ (Dressage GP Freestyle) to $0.98$ (artistic gymnastics). Each figure also shows the weighted root-mean-square deviation (RMSD) quantifying the average discrepancy between the approximated deviation and the measured values. The RMSD depends of the scale of the marking range and cannot be compared between different sports. 

With the notable exception of dressage, all sports exhibit the same heteroscedastic pattern: panel judges disagree the most when evaluating mediocre performances, and their judging error decreases as the performance quality improves towards perfection. The behavior for the worst performances depends on the sport. On the one hand, sports such as diving~(Figure~\ref{fig:Res_Diving}), trampoline (Figure~\ref{fig:Res_Trampoline}) and snowboard (Figures \ref{fig:Res_Halfpipe} and \ref{fig:Res_Slopestyle}) have many aborted or missed routines (splashing the water, stepping outside the trampoline boundaries after a jump, falling during the run). These routines result in very low marks, and the concave parabola is clearly visible for these sports, indicating smaller judging variability for performances close to zero. Smaller variability for atrocious and outstanding performances is not surprising: they either contain less components to evaluate or less errors to deduct. In both cases, this decreases the number of potential judging errors, as opposed to performances in the middle of the scoring range. On the other hand, gymnastics performances (Figures \ref{fig:Res_Acrobatics}-\ref{fig:Res_Rhythmics}) and artistic swimming routines (Figure~\ref{fig:Res_Synchronisedswimming}) barely receive a score in the lower half of the possible interval. Without these bad performances close to the minimum possible score, the quadratic fit $\hat{\sigma}_d(c)$ does not decrease towards the left border of the scoring range and can even be slightly convex.


Aerials is of particular interest because it exhibits both possible behaviors for the worst performances, which is not obvious from Figure~\ref{fig:Res_Aerials}. More precisely, the total score in aerials is the combined sum of three independent components: 'Air', 'Form' \& 'Landing'. Even though athletes do often fall when landing, this only influences their 'Landing' score and not the other two components. Figure~\ref{fig:Res_Aerialssplit} shows the aerials judging errors split per component\footnote{Some competitions in our dataset are not split per component, thus we excluded them from Figure~\ref{fig:Res_Aerialssplit}.}. The variability of the 'Landing scores', which are evenly distributed among the possible scoring range, closely follows the concave parabola, whereas the 'Air' and 'Form' components have right skewed distributions because low marks are rarely given. For these two components the quadratic regression is closer to what we observe in gymnastics or figure skating. Aerials shows at the component level what we observe at the sport level: the shape of the parabola depends on the presence or absence of performances whose quality is close to zero.

\subsection{The special case of dressage}
\begin{figure}
	\centering	\includegraphics[width=1\columnwidth]{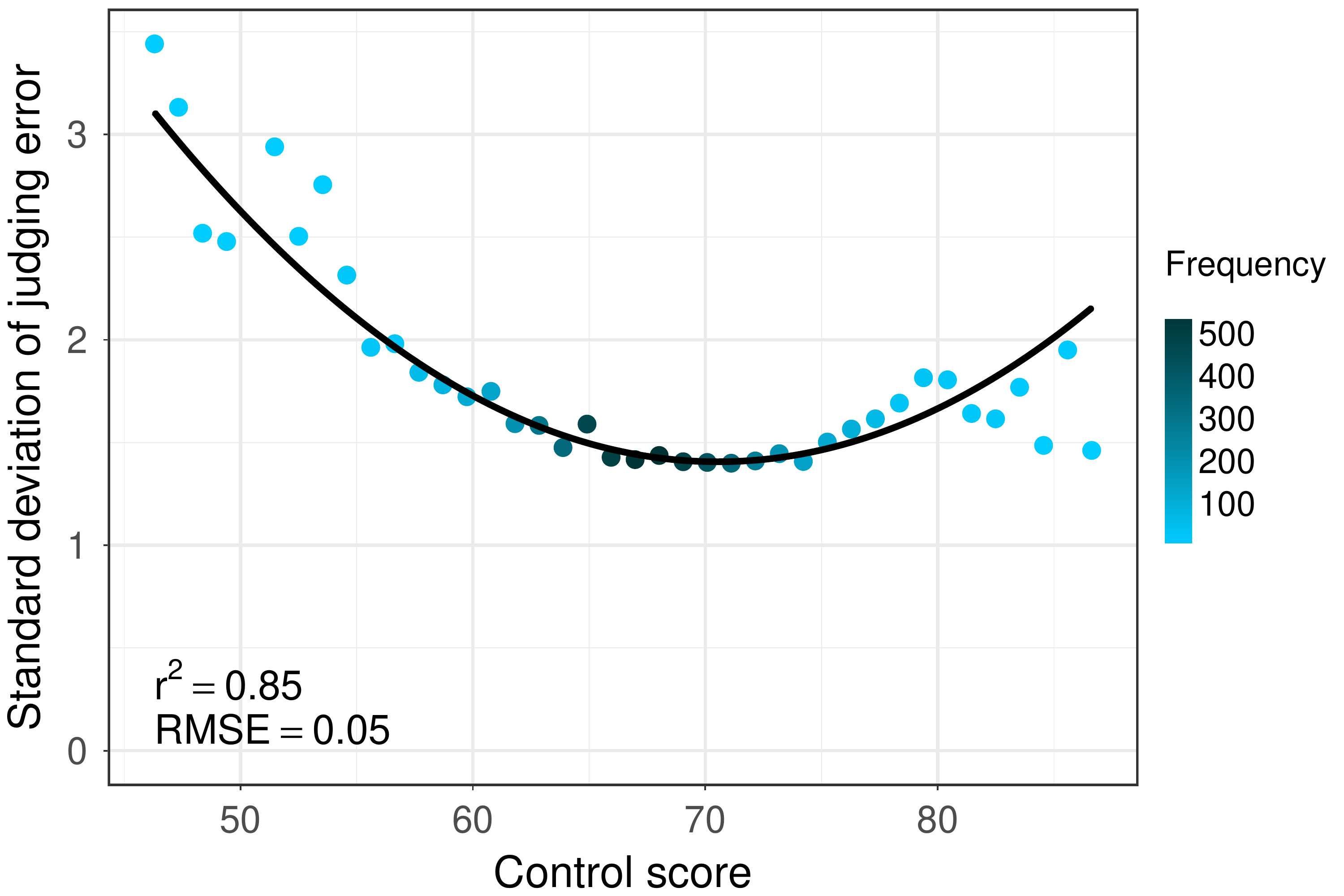}
	\caption{Standard deviation of judging marks versus performance quality in dressage GP and GP Special events.}
	\label{fig:Res_Dressage}
\end{figure}
\begin{figure}
	\centering	\includegraphics[width=1\columnwidth]{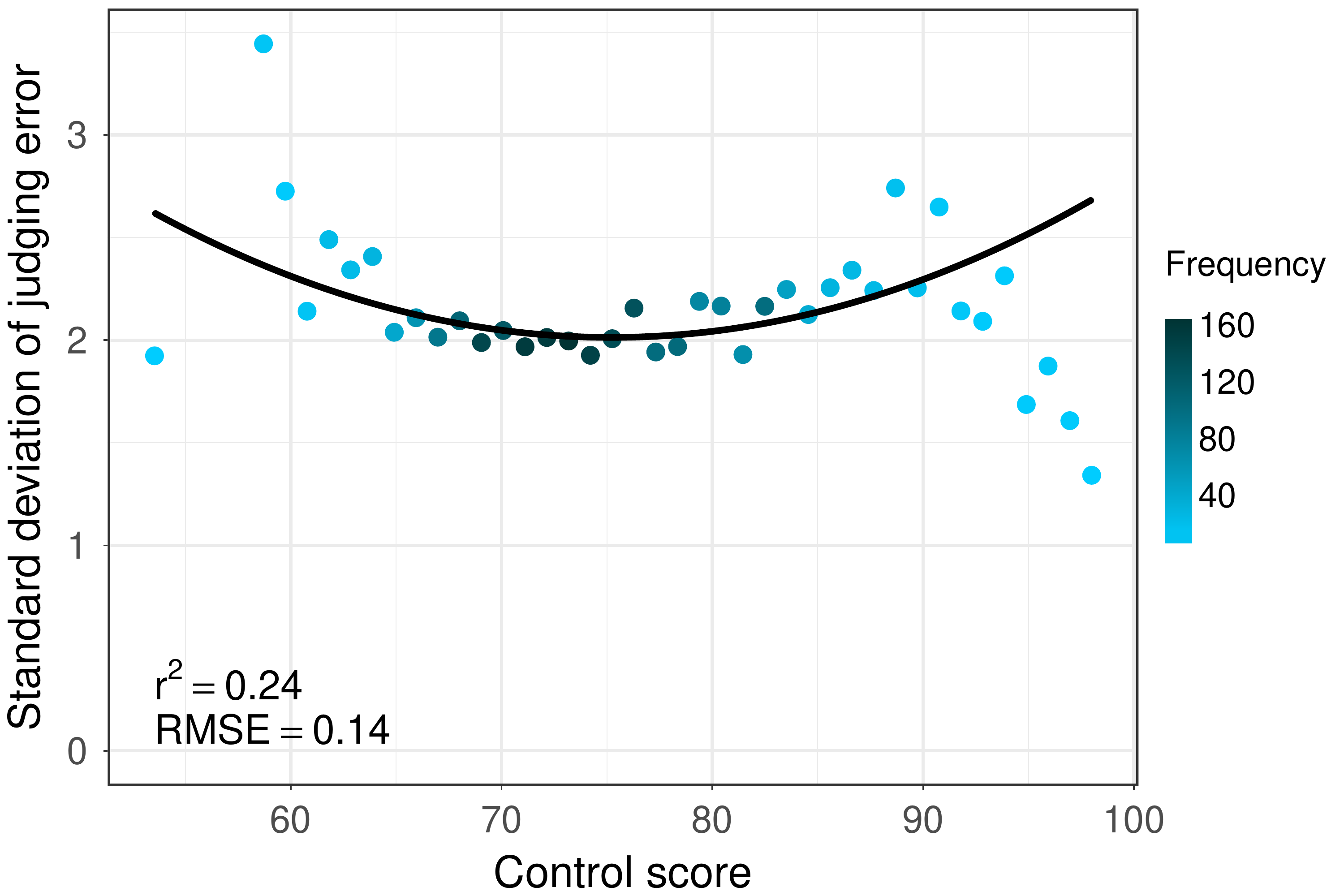}
	\caption{Standard deviation of judging marks versus performance quality in dressage split for 'Artistic presentation' scores of 'GP Freestyle to music' events.}
	\label{fig:Res_Dressage_split}
\end{figure}

Surprisingly, the observed general pattern of heteroscedasticity does not apply to dressage. Figure~\ref{fig:Res_Dressage} shows the results for standard dressage GP and GP Special competitions, whereas  Figure~\ref{fig:Res_Dressage_split} shows the results for 'GP Freestyle to music' events.

In both figures, judging errors are the lowest for average performances and the parabola is convex. For standard events in Figure~\ref{fig:Res_Dressage}, we first observe that the judges increasingly disagree as the performance quality decreases. This is similar to what we observe in gymnastics and artistic swimming and due to the fact that there are no easy to judge performances close to the lower boundary of the marking range (in dressage the lowest possible score is 0 and there is no control score below 45 in our dataset). However, judges also increasingly disagree as the performance quality approaches perfection, which is extraordinary. The judging behavior in 'GP Freestyle to music' events in Figure~\ref{fig:Res_Dressage_split} is similar but less pronounced. Furthermore, there are a few exceptional performances for which the judging errors decreases again, although this might be due to a mathematical truism: when the median mark is almost perfect, at least half the panel marks must be between the median and maximal scores, hence also perfect or close to perfect. 

We did additional analyses to understand this unexpected behavior, and found that it appears at all levels of competition in our dataset, and for every judging position around the arena. The Fédération Équestre Internationale (FEI) states that ``Dressage is the ultimate expression of horse training and elegance. Often compared to ballet, the intense connection between both human and equine athletes is a thing of beauty to behold.''\footnote{From www.fei.org/disciplines/dressage.} Elegance and beauty are highly subjective, and the subjectivity of dressage judges is not new (consult, for instance, \textcite{HMM2010}). The simplest explanation is that judges fundamentally disagree on what constitutes an above average dressage performance. This might be due to imprecise or overly subjective judging guidelines, or to the difficulty or unwillingness of judges to apply said guidelines objectively. No matter the reason, our analysis reveals a clear and systemic judging problem in dressage, and we recommend that the FEI thoroughly reviews its judging practices. We shared our results with the FEI, the International Dressage Officials Club (IDOC) and Global Dressage Analytics, who observed a similar convex parabola at the figure level\footnote{Our dataset only included the total scores, and not the individual marks per figure.}. Judges do not agree on what is, say, a 9.0 pirouette, and these disagreements are compounded over many figures, affecting at the overall performance evaluation. The FEI is currently considering major changes such as new guidelines, more precise codes of points and additional training for its judges \cite{FEI:2018}.

\section{From heteroscedasticity model to judging the judges}

The knowledge of the heteroscedastic intrinsic judging error variability $\sigma_d(c)$ makes it possible for federations to evaluate the accuracy of their judges for all the sports we analyze in this article, exactly as was done in gymnastics~\cite{MH2018:gymnastics}. We note that in practice it is often better to use weighted least-squares exponential regressions instead quadratic ones since they are more accurate for the best performances~\cite{MH2018:gymnastics}.

The marking score of judge $j$ for performance $p$ is given by $m_{p,j}\triangleq\frac{\hat{e}_{p,j}}{\hat{\sigma}(c_p)}$. This scales the error of the judge for a specific performance as a fraction of the estimated intrinsic judging error variability of the judging error for a specific discipline $d$ and performance quality $c_p$, and allows to compare judging errors for different quality levels and disciplines in an unbiased fashion. The overall marking score $M_j$ of judge $j$ is the mean squared error of all his/her judging errors in the dataset, i.e., $$M_j\triangleq \sqrt{E[m_{p,j}^2]}.$$

We can calculate the marking score of a judge for a specific competition, or longitudinally for multiple competitions over many years. A judge always marking one standard deviation $\hat{\sigma}_d(c_p)$  away from the median has a marking score of 1, and a perfect judge has a marking score of 0. The higher the marking score $M_j$, the more a judge misjudges performances compared to his/her peers. Conversely, judges with low marking scores have low noise levels around the objective performance quality.

We can use the marking score to detect outlier misjudgments, for instance judging errors greater than $2 \cdot \hat{\sigma}_d(c_p) \cdot M_j$. This flags $\approx 5\%$ of the evaluations and adjusts the threshold based on the intrinsic error variability of each judge: an accurate judge has a lower outlier detection threshold than an erratic judge. This is important to differentiate erratic but honest judges from accurate but sometimes highly biased judges. However, we must note that when using the median as the control score, a bad marking score for a single performance is not necessarily a judging error but can also mean that the judge is out of consensus. A more accurate control score is necessary to remove the ambiguity. Finally, we can also integrate the approximated standard deviation $\hat{\sigma}_d(c_p)$ and the marking score $M_j$ in bias analyses, as we did in our study of national bias in gymnastics~\cite{HM2018:NationalBias}.

\section{Conclusion}

In this article, we study judging practices of sports for which performances are evaluated by panels of judges within a finite interval. Besides the ideal of having fair competitions, these sports and the athletes that practice them have strong economic incentives to have fair and accurate judges.

We model the judging error using heteroscedastic random variables, which allows to monitor judges, detect biases, and identify outlier evaluations. With the exception of dressage, consensus among judges increases with the quality level of the performance, and we can approximate the standard deviation of the judging error accurately using a quadratic equation. Our analysis of dressage judges further shows that they increasingly disagree as the performance level increases, indicating a significant amount of subjectivity in the judging process compared to other sports with similar judging systems.

Estimating and modeling the intrinsic heteroscedasticity could also be used for other judging processes within a finite scale such as the evaluation of wine, movie and music critics. Although in these instances there is no clear notion of control score indicating the true quality level, an analysis similar to ours could quantify and highlight judges that are out of consensus with others.

\section*{Acknowledgements}

We would like to thank Nicolas Buompane, Steve Butcher, André Gueisbuhler, Sylvie Martinet and Rui Vinagre from the FIG, and Christophe Pittet and Pascal Rossier from Longines, for their help during our prior work on gymnastics that led to this extended framework. We would also like to thank Hans  Christian Matthiesen from the IDOC, Bettina De Rham from the FEI, and David Stickland from Global Dressage Analytics for discussions on our surprising observations in dressage.

\printbibliography

\end{document}